\documentclass[letterpaper,twocolumn,10pt]{article}
\usepackage[hyphens]{url}
\usepackage{usenix,epsfig,hyperref,caption,subcaption}
\hypersetup{
    bookmarksnumbered=true,
    unicode=false,          
    pdftoolbar=true,        
    pdfmenubar=true,        
    pdffitwindow=false,     
    pdfstartview={FitH},    
    pdftitle={The Anatomy of Web Censorship in Pakistan},    
    pdfnewwindow=true,      
    colorlinks=false,       
    linkcolor=red,          
    citecolor=green,        
    filecolor=magenta,      
    urlcolor=cyan           
}
\usepackage[compact]{titlesec}
\titlespacing{\section}{0pt}{2ex}{1ex}
\titlespacing{\subsection}{0pt}{1ex}{1ex}
\titlespacing{\subsubsection}{0pt}{1ex}{1ex}
\usepackage{anysize}
\marginsize{2.4cm}{2.4cm}{1.5cm}{1.0cm}
\begin{document}

\date{}

\title{\Large \bf The Anatomy of Web Censorship in Pakistan\thanks{\textbf{To
appear in: Proceedings of the 3rd USENIX Workshop on Free and Open
Communications on the Internet (FOCI '13), Washington DC, August 2013.}}}

\author{
{\rm Zubair Nabi}\\
Information Technology University, Pakistan\\
zubair.nabi@itu.edu.pk
}

\maketitle


\thispagestyle{empty}

\begin{abstract}
Over the years, the Internet has democratized the flow of information.
Unfortunately, in parallel, authoritarian regimes and other entities (such as
ISPs) for their vested interests have curtailed this flow by partially or fully
censoring the web. The policy, mechanism, and extent of this censorship varies
from country to country.

We present the first study of the cause, effect, and mechanism of web censorship
in Pakistan. Specifically, we use a publicly available list of blocked websites
and check their accessibility from multiple networks within the country. Our
results indicate that the censorship mechanism varies across websites: some are
blocked at the DNS level while others at the HTTP level. Interestingly, the
government shifted to a centralized, Internet exchange level censorship system
during the course of our study, enabling our findings to compare two generations
of blocking systems. Furthermore, we report the outcome of a controlled survey
to ascertain the mechanisms that are being actively employed by people to
circumvent censorship. Finally, we discuss some simple but surprisingly
unexplored methods of bypassing restrictions.
\end{abstract}

\section{Introduction}
In recent years, the Internet has faced an onslaught of censorship and
restrictions with local~\cite{Verkamp:2012:IMO,Mathrani:2010:WBA} and
global~\cite{Anonymous:2012:CDI} ramifications. The world over, authoritarian
regimes under the pretext of maintaining public order have been blocking web
access. This is more enunciated in the developing world where freedom of speech
and freedom of information are largely undefined. In the same vein, Pakistan has
become a poster-child for web censorship rooted in religion, politics, and
conflict/security~\cite{ONI:2012:Pakistan}. It has also been revealed as one of
the 36 countries which host FinFisher Command \& Control
servers to spy on their citizens~\cite{Boire:2013:FTE}.

According to a 2012 World Bank study, 9\% or around 16 million Pakistanis have
access to the Internet~\cite{worldbank}. Out of these 16 million users, 64\%
employ the Internet to access news websites~\cite{YG:2011:IUI}.
Therefore, the government has a high incentive to stifle this access.
Practically, filtering in Pakistan is largely geared towards blocking content
which is considered a threat to national security and/or content which is
blasphemous. The largest Internet Exchange Point (IXP) in the country is owned
by the state which simplifies the enforcement of state-wide censorship. This
censorship has been applied in waves during the past one decade---often
mandated by the judiciary~\cite{ONI:2012:Pakistan}. The side-effects of which at
times have had global impact. For instance in 2008, a naive attempt to censor
YouTube by the authorities, rendered the website unreachable for a large number
of ASes across the world~\cite{Brown:2010:PHY}.

A large number of studies have recently been conducted to study the mechanism
and effect of censorship around the world. Verkamp and
Gupta~\cite{Verkamp:2012:IMO} used PlanetLab nodes and volunteer machines to
study the mechanisms of censorship in 11 countries. One of their key insights is
that these mechanisms vary from country to country. Similarly, Mathrani and
Alipour~\cite{Mathrani:2010:WBA} presented the results of tests conducted across
10 countries using private VPNs and volunteer nodes. Their results show that
restrictions in these countries are applicable to all categories of websites:
politics, social networking, culture, news, entertainment, and religion.
Likewise, Dainotti \emph{et al.}~\cite{Dainotti:2011:ACI} made use of publicly
accessible datasets to dissect the Internet outages in Libya and Egypt during
the Arab Spring. In the same vein, a large body of
work~\cite{Xu:2011:ICC,Clayton:2006:IGF,Crandall:2007:CWT,Anonymous:2012:CDI} is
dedicated to analyzing the \emph{modus operandi} and consequence of censorship
due to the Great Firewall of China.

To the best of our knowledge, this is the first work to dissect in detail the
mechanism and effect of censorship in Pakistan. Unlike studies conducted for
other countries which relied on volunteer machines and PlanetLab nodes, we
directly use 5 different networks within Pakistan as vantage points to carry out
our tests. More importantly, during the course of our tests the country
underwent an upgrade to a central and standardized censorship system, reportedly
developed by the Canadian firm Netsweeper Inc.\footnote{The same firm has in the
past provided its filtering services to Qatar, the United Arab Emirates, Kuwait,
and Yemen~\cite{Citizen:2013:OPW}.}~\cite{Citizen:2013:OPW}.
Therefore, our results juxtapose two generations of systems. Moreover, we
present the outcome of a controlled survey to gauge the mechanisms through which
citizens are currently circumventing online blockages. Finally, we augment
these mechanisms by discussing the use of CDNs and search engine caches. Our
results can be summarized as follows:
\begin{itemize}
  \item A large number of websites are blocked using DNS injection
  \item The alternative mode of censorship in the previous system (at the ISP
  level) was HTTP 302 redirection and in case of the current system (at the
  IXP level), it is fake HTTP response injection
  \item Websites restricted at the DNS-level are also blocked at the HTTP-level
  \item A large fraction of people either use public VPN services or web proxies
  to access restricted content
  \item CDNs and search engine caches are currently viable options to access
  blocked content
\end{itemize}

The rest of the paper is organized as follows. We give a brief history of
Internet censorship in Pakistan in \S\ref{sec:background}. The methodology
employed for our study is discussed in \S\ref{sec:method}. \S\ref{sec:results}
presents the results of our tests and survey. Alternative anti-censorship
mechanisms are discussed in \S\ref{sec:discussion}. We finally conclude in
\S\ref{sec:concl} and also discuss future directions.

\section{Background}\label{sec:background}
Both telephony and Internet services in Pakistan are managed by an arm of the
state called the Pakistan Telecommunication Authority (PTA). It is in charge of
regulation and licensing of fixed-line telephony, cellular services, cable TV,
and Internet services within the country. Internet censorship is also enforced
by the government through the PTA. To give the reader some perspective,
the following is a timeline of censorship enforced by the government:
\begin{itemize}
  \item \textbf{2006: 12 websites blocked for hosting blasphemous content}. The
  content which was deemed offensive included a Blogspot blog. Lacking the
  infrastructure to block a particular blog, the entire Blogspot website was
  blocked for two months.
  \item \textbf{2008: A number of YouTube videos marked as offensive by the
  government}. Instead of implementing a URL/IP-specific restriction, an IP-wide
  block of YouTube via BGP misconfiguration was enforced, making YouTube
  inaccessible for much of the Internet for 2 hours~\cite{Brown:2010:PHY}.
  \item \textbf{2010: Facebook, YouTube, Flickr, and Wikipedia partially or
  fully blocked in reaction to ``Everybody Draw Muhammad Day''}. These websites
  were subsequently unblocked. The same year, the government also sanctioned the
  PTA to ``order temporary or permanent termination of telecom services of any
  service provider, in any part or whole of Pakistan''~\cite{Attaa:2010:GRJ}.
  \item \textbf{2012 (March): The government requests proposals for a
  country-wide URL filtering and blocking system}~\cite{ICT:2012:RFP}.
  According to the advertisement, filtering at the time was enabled by manual
  mechanisms deployed at the ISP level and the desired system was required to
  enable centralized blocking at the national IP backbone. Some other
  features\footnote{This is the first time a government has made the exact
  requirements and details of a full-fledged censorship system public.} of the
  system included:
  \begin{itemize}
    \item Filtering from domain level to sub-folder level as well as blocking of
    individual files and file types
    \item Blocking individual IPs and/or an entire range
    \item Remote network monitoring via SNMP and configuration through HTTP and
    HTTPS
    \item Operation at L2 and L3
    \item Modularity and scalability through stand-alone, plug-and-play hardware
    units capable of blocking up to 50 million URLs with a processing latency of
    less than 1ms
    \item Decoupling of policy and mechanism via storage of blacklists in an
    external database
  \end{itemize}
  \item \textbf{2012 (September): Indefinite ban on YouTube imposed in
  retaliation to a controversial movie}~\cite{IOC:2012:PYB}. The side-effects of
  this ban disrupted other Google services such as Maps, Drive, Play Store, and
  Analytics~\cite{Attaa:2012:GAM}. This was due to the fact that the same IPs
  are shared across all of these services.
\end{itemize}

\section{Methodology}\label{sec:method}
We use a publicly available dataset of
websites\footnote{\url{http://propakistani.pk/wp-content/uploads/2010/05/blocked.html}}
to perform connectivity tests. The prime reason for using this particular
dataset is that the same list of 597 websites was circulated by the PTA to all
ISPs for filtering~\cite{Attaa:2010:LOB}. While not an exhaustive list, it
provides a fairly rich set of both complete domains and subdomains for analysis.

As the list was compiled in 2010, the status of a number of websites has
changed. For instance, in 2010 the government banned a small number of YouTube
videos so the list contains individual entries for each video. Since then, the
government has blocked YouTube entirely. Therefore we removed individual video
URLs and added a single entry for YouTube. This reduced the size of the dataset
to 562 links. In addition, the list also contains a number of duplicate entries.
The removal of these entries further reduced the size of the list to 429.
Finally, a number of websites have now gone offline, mostly proxy sites,
bringing down the final tally of test sites to 307. To ensure that these final
websites are actually restricted, we tested their connectivity using a public
VPN service which terminates in the US. The results indicate that they are
accessible via VPN and are thus restricted within Pakistan.

\begin{table}[t]
\centering
    \begin{tabular}{| l | l | l |}
    \hline
    \textbf{ID} & \textbf{Nature} & \textbf{Location}\\ 
	\hline
    \emph{Network1} & University & Lahore \\
    \emph{Network2} & University & Lahore \\
    \emph{Network3} & Home & Lahore \\
    \emph{Network4} & Home & Islamabad \\
    \emph{Network5} & Cellular (EDGE) & Islamabad \\
    \hline
    \end{tabular}
    \caption{Details of Test Networks}
    \label{tab:networks}
    \vspace{-5pt}
\end{table}

\subsection{Test Script}
Our test script, dubbed \emph{Samizdat}\footnote{Available online:
\url{https://github.com/ZubairNabi/Samizdat}}, closely mimics the
CensMon~\cite{Sfakianakis:2011:CAW} system with a few modifications. Unlike
CensMon, our script does not relay test results to a remote server but logs them
locally. Furthermore, it also tries to perform DNS resolution using public
servers in addition to network-local resolvers. Finally, instead of using a
custom server to determine URL keyword filtering, it uses a public portal.

The script first downloads the list of websites and carries out the cleaning
phase described above. Then for each website it executes the following tasks:
\begin{enumerate}
  \item Performs a DNS lookup and notes the returned IP address(es). If the
  lookup fails, the website is marked as blocked and the returned DNS response,
  such as \texttt{NXDOMAIN}, \texttt{Timeout}, etc. is logged.
  The lookup is performed for both the local ISP DNS service as well as a number
  of public DNS servers: Google (\texttt{8.8.8.8}), Comodo
  (\texttt{8.26.56.26}), OpenDNS (\texttt{208.67.222.222}), Level3
  (\texttt{209.244.0.3}), and Norton (\texttt{198.153.192.40}).
  \item If the DNS lookup is successful, it tries to open a TCP socket to the IP
  address(es) on port 80 to check for IP address blacklisting. Any connection
  failures are logged.
  \item The next step is to check for URL keyword filtering. To this end, the
  URL of the website is appended to a scalable and publicly accessible portal:
  \texttt{http://www.google.com}. The script expects a HTTP \texttt{404 Not
  Found} error under normal operation. If a different code is received, the
  website is marked for URL keyword filtering.
  \item The final step is to send an HTTP request to the website. The response,
  along with its code is logged. In addition, in case of response code
  \texttt{301} or \texttt{302}, the redirection URL (from the \texttt{Location}
  field) is also logged.
\end{enumerate}
Additionally, all transient connectivity errors, such as timeouts, etc. are
logged to a separate error log.

\subsection{Test Networks}
Tests were conducted across 5 different networks: \emph{Network1-5}.
Table~\ref{tab:networks} gives the details of each network. \emph{Network1} and
\emph{Network2}, due to their academic nature, are connected to 3 and 2 ISPs,
respectively, with gigabit connectivity and auto-failover. \emph{Network3} and
\emph{Network4} on the other hand are connected to single but distinct ISPs.
\emph{Network5}, which was only used for post-April testing, is a cellular
network with GPRS-EDGE Internet connectivity. The test script was executed on
hosts within these networks at night time to minimize interaction with regular
traffic and also to minimize false positives. In addition, the measurements were
performed multiple times and on separate occasions to ensure precision. Finally,
only those results are being reported for which the error log had no entries,
i.e. there were no transient connectivity issues.

\section{Results}\label{sec:results}
In this section, we first present the results of the tests that we conducted to
investigate the mechanism and extent of censorship in the country
(\S\ref{sub:pre} and \S\ref{sub:post}) followed by the outcome of our public
survey (\S\ref{sub:survey}).

\begin{table}[t]
\centering
    \begin{tabular}{| l | l | l |}
    \hline
    \textbf{Mechanism} & \textbf{No. of Affected Sites} &
    \textbf{Percent}\\
	\hline
    DNS & 187 & 60.91\\
    IP & 0 & 0 \\
    URL-keyword & 0 & 0 \\
    HTTP (302) & 5 & 1.62 \\
    \hline
    Total & 192 & 62.53 \\
    \hline
    \end{tabular}
    \caption{Breakdown of Pre-April Test Results}
    \label{tab:pre-april}
    \vspace{-5pt}
\end{table}

\subsection{Pre-April 2013}\label{sub:pre}
Table~\ref{tab:pre-april} summarizes the results for all networks. We discuss
each of the mechanisms next.

\begin{figure*}
\begin{minipage}[t]{1\linewidth}
  \vspace*{\fill}
  \centering
  \includegraphics[width=0.9\linewidth]{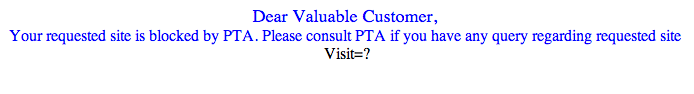}
  \vspace{-20pt}
  \subcaption{ISP 1}
  \label{fig:old}
  \includegraphics[width=1.1\linewidth]{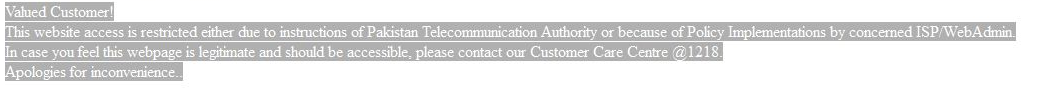}
  \vspace{-20pt}
  \subcaption{ISP 2}
  \label{fig:new}
\end{minipage}
\vspace{-5pt}
\caption{Screenshots of ISP-level Censorship Messages}
\label{fig:test}
\vspace{-5pt}
\end{figure*}

\begin{table*}[t]
\centering
    \begin{tabular}{| l | l | l | l |}
    \hline
    \textbf{No.} & \textbf{Source IP} &
    \textbf{Destination IP} & \textbf{Payload}\\
	\hline
    1. & 192.168.15.3 & 173.194.43.111 & \texttt{GET / HTTP/1.1}\\
    2. & 173.194.43.111 & 192.168.15.3 & \texttt{HTTP/1.1 302 Found
    (text/html)}\\
    3. & 192.168.15.3 & 10.16.6.41 & \texttt{GET
    /redirect.php?n=110.39.241.94@Isb-Dhok-P2\&s=124 HTTP/1.1}\\
    4. & 10.16.6.41 & 192.168.15.3 & \texttt{HTTP/1.1 200 OK (text/html)} \\
    5. & 192.168.15.3 & 10.16.6.41 & \texttt{GET /favicon.ico HTTP/1.1} \\
    6. & 10.16.6.41 & 192.168.15.3 & \texttt{HTTP/1.1 404 Not Found (text/html)}
    \\
    \hline
    \end{tabular}
    \caption{HTTP Packet-level Trace for YouTube, Pre-April}
    \label{tab:302trace}
    \vspace{-5pt}
\end{table*}

\begin{figure*}[t]
  \centering
  \includegraphics[width=1\linewidth]{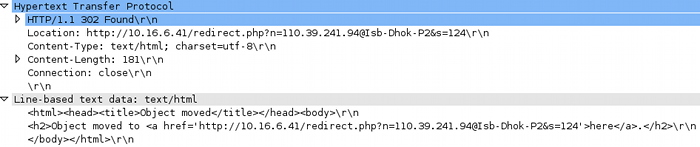}
  \vspace{-15pt}
  \caption{HTTP 302 Redirect Response for YouTube, Pre-April}
  \label{fig:302redirect}
  \vspace{-5pt}
\end{figure*}

\subsubsection{DNS}
Most websites are blocked at the DNS level. Instead of returning the proper
\texttt{A} record, a spoofed ``Non-Existent Domain'' (\texttt{NXDOMAIN}) packet
is injected and the original \texttt{A} record is suppressed, i.e. the actual
query response is never received by the client. Unlike other countries, where a
warning page is displayed~\cite{Verkamp:2012:IMO}, no such page is displayed in
this case, giving the user the impression that the page actually does not exist.
To gauge if this DNS injection is confined to resolvers within the country, the
test script also attempts to retrieve the \texttt{A} record from 5 public DNS
servers. A spoofed \texttt{NXDOMAIN} packet was observed for Google Public DNS
and Level3 as well, which suggests that DNS queries are hijacked for both
network-local and public servers. Interestingly, in case of Norton DNS, Comodo
SecureDNS, and OpenDNS, the DNS resolves to the respective service's
\texttt{NXDOMAIN} redirector, \texttt{198.153.192.3}, \texttt{92.242.144.50},
and \texttt{67.215.65.132}, respectively. This suggests that the censoring
system is mindful of the behaviour of individual resolvers.

\subsubsection{IP and URL-keyword Filtering}
Post-DNS resolution, the script was able to connect to the IPs of all websites
on port 80, showing that no IP-level blacklisting is in place. Similarly, the
script received a \texttt{404 Not Found} for each website after appending its
URL to \texttt{www.google.com}, indicating that URL keyword filtering is also
not in effect.

\subsubsection{HTTP}
A large number of websites exhibit normal behaviour and return either a 200
(\texttt{OK}) or 203 (\texttt{Non-Authoritative Information}) code. A few sites
also have legitimate 301 (\texttt{Moved Permanently}) or 302 (\texttt{Found})
redirects, for instance from \texttt{http://www.anonymizer.com} to
\texttt{https://www.anonymizer.com/index.html}. In total, 115 websites exhibited
normal behaviour and have thus been unblocked by the authorities since the time
the dataset was compiled.

5 websites, including YouTube, result in redirection to a warning screen. The
warning screen varies from ISP to ISP (shown in Figure~\ref{fig:test}) which
indicates that this redirection is done at the ISP-level.
Table~\ref{tab:302trace} presents the HTTP packet level trace\footnote{Captured
using Wireshark (\url{http://www.wireshark.org/}).} for YouTube. It is clear
from the table that the HTTP \texttt{GET} request is intercepted and the session
is redirected to a private IP which displays the warning page.
Figure~\ref{fig:302redirect} shows the HTTP 302 redirect response packet in
which the \texttt{Location} field contains the redirection IP for one particular
ISP. The same sequence of packets, although with different redirection IPs, was
observed on all 5 networks. This censorship is triggered by a combination of the
hostname and object URI (\texttt{Host} and \texttt{Request URI} fields) within
the HTTP request header. We ascertained this by creating a custom HTTP request
header for all 5 websites and sending it to an uncensored URL instead of the
actual one. This resulted in redirection to the warning page in each case. For
instance, setting the hostname as \texttt{youtube.com} triggers censorship
regardless of the destination URL. In contrast, the hostname \texttt{vimeo.com}
is allowed to go through but the hostname in conjunction with the URI
\texttt{64414932} experiences redirection. It is also important to highlight
that this censorship is an example of ``filter and return'' as opposed to
``allow but return first''~\cite{Verkamp:2012:IMO}, i.e. the original session is
disrupted and suppressed by the censoring module.

\subsection{Post-April 2013}\label{sub:post}
\begin{figure}[t]
\centering
  \includegraphics[width=1\linewidth]{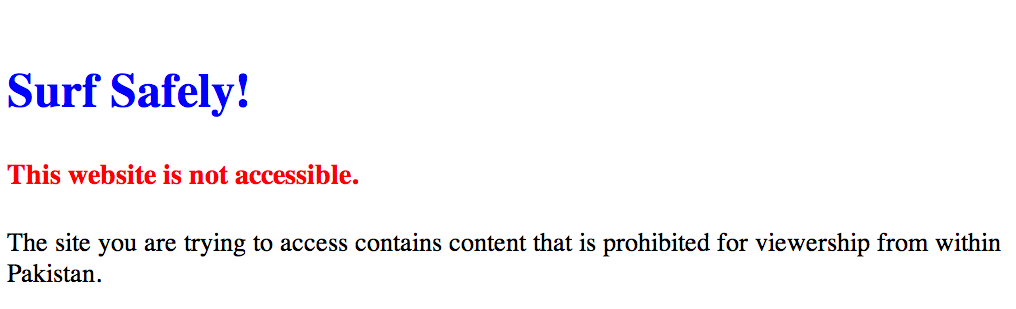}
  \vspace{-20pt}
  \caption{Censorship Warning Page, Post-April 2013}
  \label{fig:postaprilscreen}
  \vspace{-5pt}
\end{figure}

\begin{figure}[t]
  \centering
  \includegraphics[width=1\linewidth]{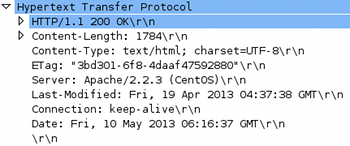}
  \vspace{-10pt}
  \caption{HTTP Response for YouTube, Post-April}
  \label{fig:200paste}
  \vspace{-5pt}
\end{figure}

In May 2013 we serendipitously noticed a new warning page for YouTube. As a
consequence, we re-conducted all tests. The results of which are summarized in
Table~\ref{tab:post-april}. The number of websites affected by DNS, IP, and
URL-keyword based filtering remained largely the same. In contrast, websites
which experienced HTTP 302 redirection are now displaying a different warning
page (shown in Figure~\ref{fig:postaprilscreen}) without any explicit
redirection. Packet level inspection revealed that the legitimate HTTP response
is being replaced with a spoofed response which displays the said page.
Table~\ref{tab:200trace} shows the sequence of HTTP packets for YouTube. Instead
of the actual response, a packet with the content of the warning page and status
code 200 is injected. The restriction system uses the first HTTP \texttt{GET}
request as a censoring trigger based on the hostname and URI, similar to the
mechanism discussed in the previous section. The status code of 200 makes the
browser believe that it is a normal response, thus restricting it from fetching
content from the intended destination (YouTube in this case). As a result, the
actual TCP connection established with YouTube will eventually timeout for lack
of activity. Figure~\ref{fig:200paste} presents the HTTP response message
received for all websites which experienced this category of censorship. The
\texttt{Last-Modified} field of the response contains the value \texttt{Fri, 19
Apr 2013} which we suspect is the date this new system came into effect.
Moreover, all of our test networks experienced the same censorship mechanism and
had the same warning page, except for \emph{Network4} which was still under the
influence of ISP-level filtering. This has two main implications: (1) The
country has moved from fragmented ISP-level to IXP-level filtering, and (2) The
transition from the old system to the new one is taking place in phases.

\begin{table}[h]
\centering
    \begin{tabular}{| l | l | l |}
    \hline
    \textbf{Mechanism} & \textbf{No. of Affected Sites} &
    \textbf{Percent}\\
	\hline
    DNS & 179 & 58.30 \\
    IP & 0 & 0 \\
    URL-keyword & 0 & 0 \\
    HTTP (200) & 5 & 1.62 \\
    \hline
    Total & 184 & 59.92 \\
    \hline
    \end{tabular}
    \caption{Breakdown of Post-April Test Results}
    \label{tab:post-april}
    \vspace{-5pt}
\end{table}

\begin{table*}[t]
\centering
    \begin{tabular}{| l | l | l | l |}
    \hline
    \textbf{No.} & \textbf{Source IP} &
    \textbf{Destination IP} & \textbf{Payload}\\
	\hline
    1. & 192.168.43.112 & 173.194.35.100 & \texttt{GET / HTTP/1.1}\\
    2. & 173.194.35.100 & 192.168.43.112 & \texttt{HTTP/1.1 200 OK (text/html)}\\
    3. & 192.168.43.112 & 173.194.35.100 & \texttt{GET /favicon.ico HTTP/1.1}\\
    4. & 173.194.35.100 & 192.168.43.112 & \texttt{HTTP/1.1 200 OK
    (image/x-icon)}
    \\
    \hline
    \end{tabular}
    \caption{HTTP Packet-level Trace for YouTube, Post-April}
    \label{tab:200trace}
    \vspace{-5pt}
\end{table*}

\subsection{Survey}\label{sub:survey}
\begin{figure}[h]
  \centering
  \includegraphics[width=1\linewidth]{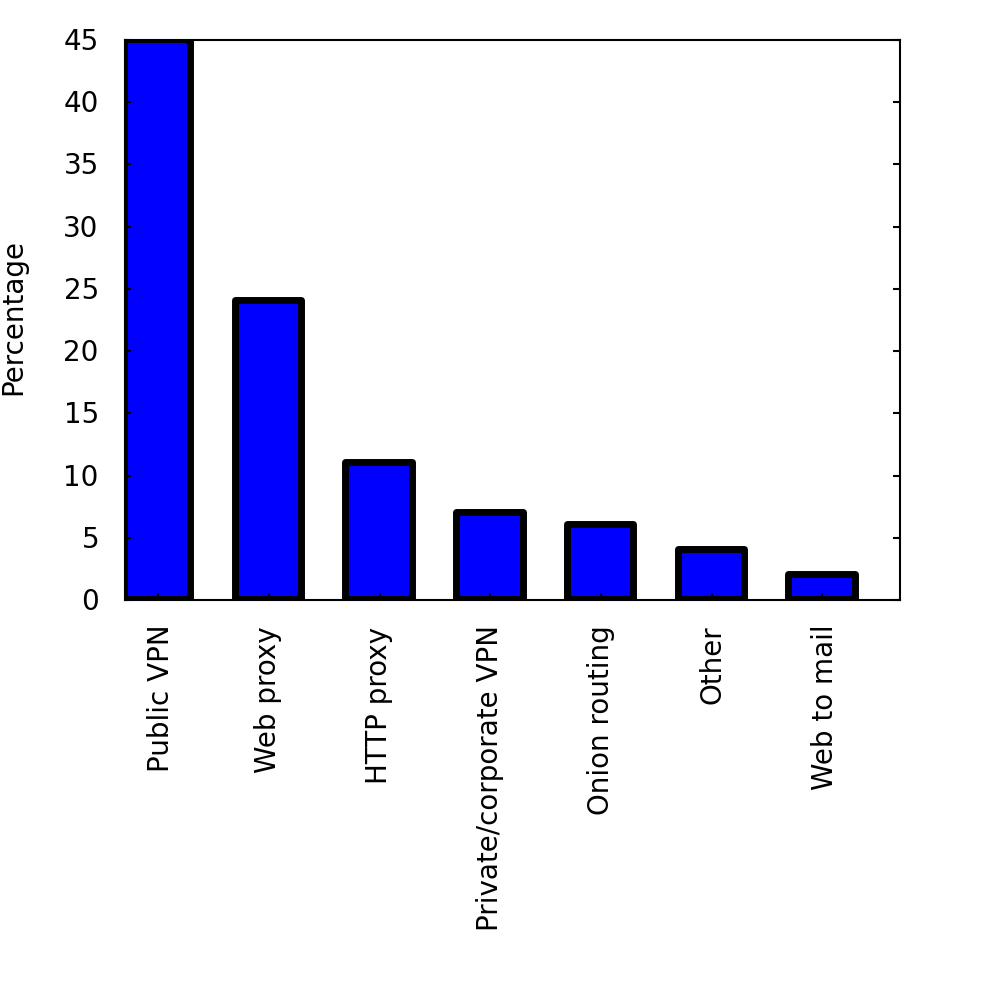}
  \vspace{-30pt}
  \caption{Survey Results}
  \label{fig:survey}
  \vspace{-5pt}
\end{figure}

Having worked out the censorship mechanisms, we conducted a controlled survey to
ascertain the tools people are using to circumvent censorship in the country.
Due to the sensitive nature of the subject matter we only shared it with
personal contacts. Figure~\ref{fig:survey} presents the results for 67
respondents, mostly fellow computer scientists. Public VPN services are
predominantly used; accounting for around 45\%. Specifically, Hotspot
Shield\footnote{\url{http://www.hotspotshield.com/}} and
Spotflux\footnote{\url{http://www.spotflux.com/}}---both of which provide free
VPN services through client applications---are popular choices. Web proxies
(24\%) and HTTP proxies (11\%), such as
Ultrasurf\footnote{\url{https://ultrasurf.us/}} also have a substantial
user-base. It is noteworthy that the respondents are more technically aware than
the average citizen and thus the results might be biased towards solutions that
require above average computer skills.

\section{Alternative Circumvention Mechanisms}\label{sec:discussion}
In this section we discuss some alternative methods of bypassing restrictions.
We first analyze web-based DNS lookup followed by content distribution networks
and search engine caches.

\subsection{Web-based DNS Lookup}
A well-known mechanism to bypass DNS-based filtering is to use a public DNS
service~\cite{Verkamp:2012:IMO}. Our findings indicate that Pakistan also
injects a fake DNS response for public DNS services, thus negating their use.
Another means to resolve hostnames to IPs is to use an online service. To check
if this was a viable option to bypass filtering, we wrote a script to, (1)
resolve hostnames to IPs using an online
service\footnote{\url{http://ping.eu/nslookup/}}, and (2) use a custom HTTP
header to fetch the webpage. The custom HTTP header is necessary as many
multi-service sites share the same IPs across many of their services. For
instance, Google uses the same set of IPs for YouTube, Drive, etc. Therefore,
these resolved IPs cannot be used as drop-in replacements for hostnames. To
remedy this, the HTTP header needs to contain the service name within its
\texttt{Host} field. Our results show that all websites which experience
DNS-based censorship, are also censored using secondary mechanisms (similar to
South Korea~\cite{Verkamp:2012:IMO}): HTTP 302 redirection in case of the
pre-April system and fake HTTP 200 response in case of the current post-April
system. This also rules out the use of user-generated
content~\cite{Burnett:2010:CAC,Invernizzi:2012:MIA} to host DNS records.
Furthermore, according to Verkamp and Gupta~\cite{Verkamp:2012:IMO}, South Korea
makes use of DNS filtering for DNS entries that resolve to a single site.
Whereas for IPs which are shared across several hostnames and the filtering
needs to be selective, the HTTP-level mechanism is used. We found a similar
pattern in our tests. For instance, multi-IP and multi-service sites such as
YouTube and Wikipedia are only blocked at the HTTP-level while others are
blocked at both levels.
 
\subsection{Content Distribution Networks}
CDNs have increased the availability, reliability, and performance of the web by
geo-caching content. To check if any of the blocked websites were accessible
through CDNs, we tried to fetch them from the peer-to-peer
CoralCDN\footnote{\url{http://www.coralcdn.org/}}. Websites are \emph{coralized}
by appending \texttt{.nyud.net} to the hostname. For instance,
\texttt{http://google.com.nyud.net} is the coralized version of
\texttt{http://google.com}. The original test script was modified to attempt to
fetch each website from CoralCDN. The results indicate that all 307 websites are
accessible through this method. Previous studies have already shown that
CoralCDN is being used to bypass censorship in some
areas~\cite{Freedman:2010:ECF}.

\subsection{Search Engine Caches}
Search engines usually cache snapshots of indexed pages. These can readily be
accessed online. In case of Google, this requires pre-pending \texttt{cache:} to
the URL. We checked the status of all 307 websites from our dataset and found
out that all of them are accessible through the Google cache. Bing and Internet
Archive\footnote{\url{http://archive.org}} also yielded similar results.

\section{Conclusion and Future Work}\label{sec:concl}
We presented the first study of the cause, effect, and mechanism of Internet
censorship in Pakistan. During the course of our work, in April 2013, the
country underwent an upgrade from ISP-level blocking to a centralized system,
allowing us to juxtapose two different generations of techniques. Our discovery
of the upgrade to IXP-level filtering is consistent with the findings of a study
conducted by the Citizen Lab~\cite{Citizen:2013:OPW}, coincidentally around the
same time. To work out the specifics of restrictions, we used a publicly
available list of 307 websites. For diversity and precision, 5 distinct networks
from within the country were employed. Our results show that DNS injection is
the predominant mechanism for blocking websites. This is applicable to both
local DNS resolvers as well as public resolvers such as Google DNS and OpenDNS.
The next line of censorship is at the HTTP-level. In case of the pre-April
scheme, HTTP 302 redirection is used to disrupt and suppress sessions and in the
post-April scheme, a fake HTTP 200 response packet is injected to give the
browser the illusion that the session has been completed. In addition, websites
blocked at the DNS-level also experience restrictions at the HTTP-level.
Furthermore, the outcome from our controlled survey shows that public VPN
services and web proxies are the two most popular tools to bypass restrictions.
Finally, we showed that CDNs and search engine caches are simple but
surprisingly unexplored means of accessing blocked content. 

This work provided an initial window into the Internet censorship regime in
Pakistan but a complete picture requires an expansion in the number of test
websites as well as networks, which constitutes our future work. Furthermore, it
is also not clear how the censoring module determines the exact
\texttt{NXDOMAIN} redirector for public DNS resolvers, i.e. whether it maintains
a list of all resolvers and their redirectors or it queries the actual resolver
to obtain its redirector upon each lookup. Additionally, our future work also
includes the examination of the side-effects of DNS injection on the same lines
as~\cite{Anonymous:2012:CDI}.

\section*{Acknowledgements}
The author would like to thank the anonymous reviewers for their comments,
concerns, and suggestions which greatly improved the quality of this
publication.

\newpage

{\footnotesize \bibliographystyle{acm}
\bibliography{foci}}

\end{document}